\documentclass[prm,twocolumn,superscriptaddress]{revtex4}
\usepackage[utf8]{inputenc}
\usepackage{xcolor}
\usepackage{amsmath}
\usepackage{graphicx}
\usepackage{booktabs}

\begin{document}

\title{Evolution of charge density waves from three-dimensional to quasi-two-dimensional in Kagome superconductors Cs(V$_{1-x}M_{x}$)$_3$Sb$_5$ ($M$ = Nb, Ta)
}
\author{Qian Xiao}\thanks{These authors contributed equally to this work.}
\affiliation{International Center for Quantum Materials, School of Physics, Peking University, Beijing 100871, China}

\author{Qizhi Li}\thanks{These authors contributed equally to this work.}
\affiliation{International Center for Quantum Materials, School of Physics, Peking University, Beijing 100871, China}

\author{Jinjin Liu}\thanks{These authors contributed equally to this work.}
\affiliation{Centre for Quantum Physics, Key Laboratory of Advanced Optoelectronic Quantum Architecture and
Measurement (MOE), School of Physics, Beijing Institute of Technology, Beijing 100081, China}
\affiliation{Beijing Key Lab of Nanophotonics and Ultrafine Optoelectronic Systems, Beijing Institute of Technology, Beijing 100081, China}

\author{Yongkai Li}
\affiliation{Centre for Quantum Physics, Key Laboratory of Advanced Optoelectronic Quantum Architecture and Measurement (MOE), School of Physics, Beijing Institute of Technology, Beijing 100081, China}
\affiliation{Beijing Key Lab of Nanophotonics and Ultrafine Optoelectronic Systems, Beijing Institute of Technology, Beijing 100081, China}
\affiliation{Material Science Center, Yangtze Delta Region Academy of Beijing Institute of Technology, Jiaxing, 314011, China}

\author{Wei Xia}
\affiliation{School of Physical Science and Technology, ShanghaiTech University, Shanghai 201210, China}
\affiliation{ShanghaiTech Laboratory for Topological Physics, ShanghaiTech University, Shanghai 201210, China}

\author{Xiquan Zheng}
\affiliation{International Center for Quantum Materials, School of Physics, Peking University, Beijing 100871, China}

\author{Yanfeng Guo}
\affiliation{School of Physical Science and Technology, ShanghaiTech University, Shanghai 201210, China}
\affiliation{ShanghaiTech Laboratory for Topological Physics, ShanghaiTech University, Shanghai 201210, China}

\author{Yugui Yao}
\affiliation{Centre for Quantum Physics, Key Laboratory of Advanced Optoelectronic Quantum Architecture and Measurement (MOE), School of Physics, Beijing Institute of Technology, Beijing 100081, China}
\affiliation{Beijing Key Lab of Nanophotonics and Ultrafine Optoelectronic Systems, Beijing Institute of Technology, Beijing 100081, China}

\author{Zhiwei Wang}
\email{zhiweiwang@bit.edu.cn}
\affiliation{Centre for Quantum Physics, Key Laboratory of Advanced Optoelectronic Quantum Architecture and
Measurement (MOE), School of Physics, Beijing Institute of Technology, Beijing 100081, China}
\affiliation{Beijing Key Lab of Nanophotonics and Ultrafine Optoelectronic Systems, Beijing Institute of Technology, Beijing 100081, China}
\affiliation{Material Science Center, Yangtze Delta Region Academy of Beijing Institute of Technology, Jiaxing
314011, China}

\author{Yingying Peng}
\email{yingying.peng@pku.edu.cn}
\affiliation{International Center for Quantum Materials, School of Physics, Peking University, Beijing 100871, China}
\affiliation{Collaborative Innovation Center of Quantum Matter, Beijing 100871, China}

\date{\today}

\begin{abstract}
The Kagome material $A{\mathrm{V}}_3{\mathrm{Sb}}_5$ ($A$ = K, Rb, Cs) with geometry frustration hosts non-trivial topological electronic structures, electronic nematicity, charge density wave (CDW) and superconductivity, providing an ideal platform to study the interplay between these phases. Specifically, in pressurized- or substituted-${\mathrm{CsV}}_3{\mathrm{Sb}}_5$, the relationship between CDW and superconductivity is unusual and remains to be fully understood. 
Recently, coexisting and competing 2 $\times$ 2 $\times$ 4 and 2 $\times$ 2 $\times$ 2 CDW phases were discovered in ${\mathrm{CsV}}_3{\mathrm{Sb}}_5$. To investigate the evolution of the CDW phases with the substitution of V atoms, we performed x-ray diffraction (XRD) experiments on ${\mathrm{Cs(V}}_{1-x}{\mathrm{Ta}}_{x}{\mathrm{)}}_3{\mathrm{Sb}}_5$ and ${\mathrm{Cs(V}}_{1-x}{\mathrm{Nb}}_{x}{\mathrm{)}}_3{\mathrm{Sb}}_5$. Our results indicate that in all substituted samples, the discrete CDW reflection points in pristine ${\mathrm{CsV}}_3{\mathrm{Sb}}_5$ change to rod-like structures along the $c^\star$ direction. This suggests that the long-ranged three-dimensional CDW becomes quasi-two-dimensional by the substitution of V by Ta/Nb. Moreover, our temperature-dependent measurements show that there is no hysteresis behavior of CDW signals, indicating that the 2 $\times$ 2 $\times$ 4 CDW phase is easily suppressed by even a slight substitution of V with Nb/Ta. These findings uncover the CDW evolution upon substitution of V atoms in CsV$_3$Sb$_5$, providing insights into the microscopic mechanism of CDW and helping to understand the interplay between intertwined phases and superconductivity.
\end{abstract}

\maketitle

\section{Introduction}
The Kagome nets, comprised of three interwoven triangular lattices, are geometrically frustrated and have been predicted to possess flat bands, Dirac cones, and van Hove singularities (vHSs) in theory, offering an opportunity for the realization of novel topological quantum states such as quantum spin liquid\,\cite{S.Yan2011science,2016RevModPhys,2017RevModPhys}. At van Hove filling, a diverse phase diagram emerges with various electronic instabilities such as spin and charge bond orders, superconductivity, and charge density wave (CDW), all of which can be controlled by Hubbard interactions $U$\,\cite{chiralSDW,instabilities.PRL,instabilities.prb}. The Kagome family $A$V$_3$Sb$_5$ ($A$ = K, Rb, Cs) has been identified as a topological metal with a superconducting transition occurring below 3 K, leading to a surge in research in condensed matter physics\,\cite{Brenden.prl,prm2021sc.KVS,cpl2021RVS.sc}, and numerous exotic physical properties of this family were revealed. A CDW transition occurs between 78 K and 103 K\,\cite{Brenden.prl,prm2021sc.KVS,cpl2021RVS.sc,jiangyx}, and abnormal Hall effect (AHE) emerges simultaneously, reaching a magnitude of 10,000 $\Omega^{-1}$cm$^{-1}$ at low temperature without any detected local spin moments\,\cite{AHE.KVS,AHE.CVS,umR.kenny}. By substituting V sites, AHE decreases with decreasing CDW, indicating a strong correlation between CDW and AHE\,\cite{Tidoped,CVNS_phaseDiagram}. Moreover, a twofold rotation symmetry is observed in the CDW state\,\cite{NLWang.TRS,TRSB.NematicDomain.natphy,HongLi.natphy} and persists into the superconducting phase in CsV$_3$Sb$_5$\,\cite{2symmetry2021natcom}. Electronic nematicity develops below CDW transition temperature $T_{\mathrm{CDW}}$, which undergoes a nematic transition at around 35 K\,\cite{nematic.CDW.nat}. Muon spin relaxation measurements suggest the concurrence of the time-reversal symmetry breaking (TRSB) and CDW phase formation\,\cite{miuSR.nat,TRSB.NematicDomain.natphy}. Thus, investigating the CDW phases is crucial for comprehending these exotic properties in $A$V$_3$Sb$_5$. 

CsV$_3$Sb$_5$ exhibits a rich array of exotic properties, including coexisting and competing 2 $\times$ 2 $\times$ 4 and 2 $\times$ 2 $\times$ 2 CDW phases at low temperature, which have been demonstrated by combining x-ray scattering experiments with density-functional theory calculations\,\cite{xq}.  
It is therefore important to investigate how external stimuli such as pressure and substitution may affect the CDW phase and other properties, especially superconductivity. Previous studies on pressurized and Sn-doped CsV$_3$Sb$_5$ reveal an unusual competition between CDW and superconductivity, where superconducting transition temperature $T_{\mathrm{c}}$ shows double peaks with the suppression of CDW phases\,\cite{F.H.Yu.pressure.nc,ChenKY.pressure.PRL,doping.2sc}. This unique behavior in CsV$_3$Sb$_5$ differs from that of other $A$V$_3$Sb$_5$ family members\,\cite{KVSSn.RVSSn.prm,review.natphy}. 
Notably, substituting V sites has been shown to have a distinct effect from doping Sb sites in CsV$_3$Sb$_5$. Doping Sb with Sn resulted in the observation of prominent double $T_{\mathrm{c}}$ domes\,\cite{doping.2sc}, while $T_{\mathrm{c}}$ entered a plateau after the initial enhancement in ${\mathrm{Cs(V}}_{1-x}{\mathrm{Ta}}_{x}{\mathrm{)}}_3{\mathrm{Sb}}_5$\,\cite{CVTS.CVNS.ARPES}. 
From an electronic structure perspective, holes are introduced in ${\mathrm{CsV}}_3{\mathrm{Sb}}_{5-x}$Sn$_x$, which lifts the Sb-derived electron-like band at $\Gamma$ point up and above Fermi level, but only causes a relatively small change in the V-derived vHS at the $M$ point\,\cite{doping.2sc}. 
In contrast, the isovalent substitution of V by Nb lifts the saddle point up away from the Fermi level and shifts the electron-like band at $\Gamma$ point downward\,\cite{CVNS.ARPES.PRL,CVNS_phaseDiagram}. Thus, substituting V sites has a distinct effect from doping Sb sites in ${\mathrm{CsV}}_3{\mathrm{Sb}}_5$.  Previous x-ray diffraction (XRD) studies on CsV$_3$Sb$_{5-x}$Sn$_x$ found that the interlayer CDW correlations become short-ranged upon doping Sb by Sn, with Sn mainly occupying the Sb sites in the Kagome plane\,\cite{CVSSn.xrd,doping.2sc}. However, XRD study on CsV$_3$Sb$_5$-derived Kagome material with V substitution is still lacking. The CDW phase, which is accompanied by the deformation of vanadium Kagome net, is widely believed to arise from the scattering between vHSs at the $M$ point\,\cite{jiangyx,CVNS_phaseDiagram,CVNS.ARPES.PRL}. Therefore, it is interesting to study the effect of substituting V sites on the CDW phase. 

Here we employ x-ray diffraction to investigate the evolution of CDW phases by substituting V with Ta and Nb in CsV$_3$Sb$_5$. As substitution levels increase, the CDW transition temperature decreases smoothly, while the enhancement of superconducting transition temperature is non-monotonic. 
Compared to pristine CsV$_3$Sb$_5$, the XRD patterns of Nb and Ta substituted CsV$_3$Sb$_5$ show rod-like CDW signals along the $c^\star$-direction, indicating that CDW correlations along the $c$-direction are rapidly suppressed by a slight substitution. The estimated correlation length along the out-of-plane direction is only $\sim$ 2c in ${\mathrm{Cs(V}}_{1-x}{\mathrm{Ta}}_{x}{\mathrm{)}}_3{\mathrm{Sb}}_5$ ($x$ = 0.067). Our temperature-dependent XRD measurements on substituted CsV$_3$Sb$_5$ samples demonstrate a complete suppression of the 2 $\times$ 2 $\times$ 4 CDW phase, evidenced by the disappearance of the hysteresis behavior of CDW signals. Furthermore, the CDW transition width broadens with increasing substitution levels. These findings shed light on understanding the microscopic mechanism of CDW and the interplay between the intertwined phases in Kagome superconductor CsV$_3$Sb$_5$.  

\section{Materials and methods}
We have studied three substitution levels of ${\mathrm{Cs(V}}_{1-x}{\mathrm{Ta}}_{x}{\mathrm{)}}_3{\mathrm{Sb}}_5$: $x$ = 0.04, 0.067, 0.1, and four substitution levels of ${\mathrm{Cs(V}}_{1-x}{\mathrm{Nb}}_{x}{\mathrm{)}}_3{\mathrm{Sb}}_5$: $x$ = 0.021, 0.023, 0.026 and 0.068. 
The crystalline ${\mathrm{CsV}}_3{\mathrm{Sb}}_5$, ${\mathrm{Cs(V}}_{1-x}{\mathrm{Ta}}_{x}{\mathrm{)}}_3{\mathrm{Sb}}_5$ and ${\mathrm{Cs(V}}_{1-x}{\mathrm{Nb}}_{x}{\mathrm{)}}_3{\mathrm{Sb}}_5$ samples in present XRD study were grown by a self-flux growth method\,\cite{xq,CVTS.CVNS.ARPES}. Chemical composition was identified on a JEOL scanning electron microscope (SEM, JSM-7500F) equipped with energy dispersive x-ray spectroscopy (EDX). Electronic transport measurements with temperatures down to 1.8 K were carried out on a physical property measurement system (PPMS, Quantum Design). The contacts were prepared by using gold wires which were attached by room-temperature-cured silver paste to the sample. 

We used a custom-designed x-ray instrument equipped with a Xenocs Genix3D Mo K$_\alpha$ (17.48 keV) x-ray source to perform single crystal x-ray diffraction measurements. The instrument provided a beam spot size of 150 $\mu$m at the sample position and $\sim$ 2.5$\times$10$^7$ photons/sec. We also performed high-resolution XRD measurements on pristine ${\mathrm{CsV}}_3{\mathrm{Sb}}_5$ and Ta-substituted ${\mathrm{CsV}}_3{\mathrm{Sb}}_5$ single crystals using a highly monochromatic Cu
K$_{\alpha1}$ (8.048 keV) radiation x-ray source with $\sim$ 4$\times$10$^6$ photons/sec\,\cite{xq}. The samples were mounted on a 4-circle Huber diffractometer and cooled by a closed-cycle cryostat with a temperature accuracy of about $\pm$ 91 mK and temperature stability of about $\pm$ 30 mK. One beryllium (Be) dome served as a vacuum and radiation shield to cool the sample down to 18 K. We collected scattering signals using a highly sensitive single-photon counting PILATUS3 R 1M solid state area detector with 980$\times$1042 pixels, each pixel size being $172~\mu m \times 172~\mu m$. To map the 3-dimensional momentum space, we took images in 0.1$^\circ$ increments while rotating the samples\,\cite{xq}. 

We indexed the x-ray diffraction pattern of Cs(V$_{1-x}M_{x}$)$_3$Sb$_5$ ($M$ = Nb, Ta) using a hexagonal unit cell based on the main Bragg peaks\,\cite{xq}. In pristine ${\mathrm{CsV}}_3{\mathrm{Sb}}_5$, the lattice parameters are $a$ = $b$ $\simeq$ 5.53 ${\mathrm{\AA}}$ and $c$ $\simeq$ 9.28${\mathrm{\AA}}$. While the $c$ lattice parameter remains almost unchanged in the Nb/Ta substituted CsV$_3$Sb$_5$, the in-plane lattice parameter $a$ (= $b$) becomes larger by substituting V with Ta, which increases to 5.57 ${\mathrm{\AA}}$ for ${\mathrm{Cs(V}}_{1-x}{\mathrm{Ta}}_{x}{\mathrm{)}}_3{\mathrm{Sb}}_5$ ($x$ = 0.1). We did not observe any discernible change in the lattice parameters for Nb-doped samples compared to the pristine samples.  
Throughout this manuscript, we label the Miller indices ($H, K, L$) according to the undistorted high-temperature phase (see details of indexation in ref.\,\cite{xq}).

\section{Results}
\subsection{Resistivity measurements}

\begin{figure}[htbp]
\centering
\includegraphics[width=\columnwidth]{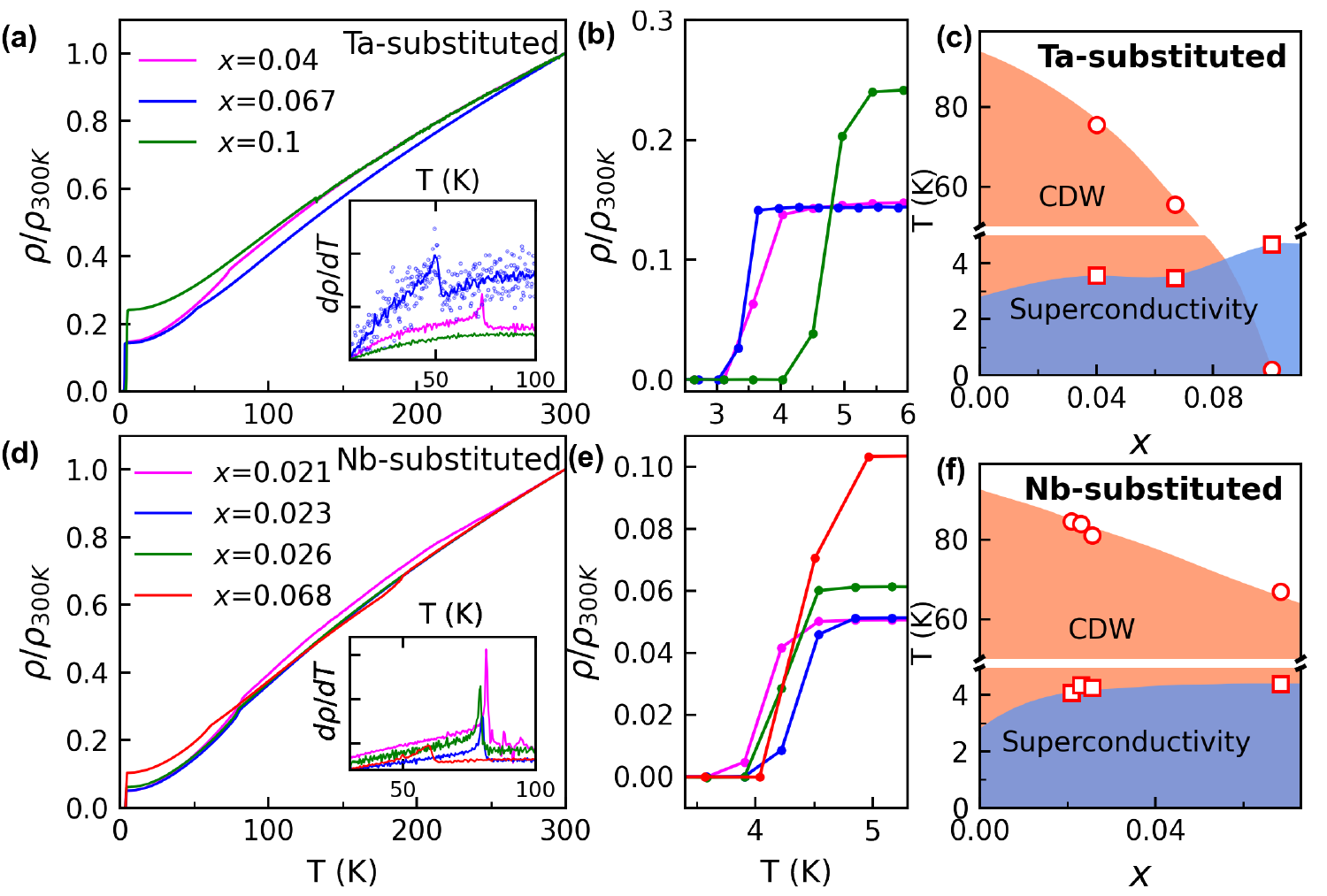}
\caption{(a)(d) Temperature dependence of in-plane resistivity measured from 1.8 K to 300 K for ${\mathrm{Cs(V}}_{1-x}{\mathrm{Ta}}_{x}{\mathrm{)}}_3{\mathrm{Sb}}_5$ and ${\mathrm{Cs(V}}_{1-x}{\mathrm{Nb}}_{x}{\mathrm{)}}_3{\mathrm{Sb}}_5$, respectively. The data were normalized to the resistivity at 300 K; the inset shows d$\rho$/dT as a function of temperature near the CDW transition. (b)(e) Zoomed-in views of the $\rho$(T) curves near the superconductivity transition temperatures for ${\mathrm{Cs(V}}_{1-x}{\mathrm{Ta}}_{x}{\mathrm{)}}_3{\mathrm{Sb}}_5$ and ${\mathrm{Cs(V}}_{1-x}{\mathrm{Nb}}_{x}{\mathrm{)}}_3{\mathrm{Sb}}_5$, respectively. The schematic phase diagrams of ${\mathrm{Cs(V}}_{1-x}{\mathrm{Ta}}_{x}{\mathrm{)}}_3{\mathrm{Sb}}_5$ and ${\mathrm{Cs(V}}_{1-x}{\mathrm{Nb}}_{x}{\mathrm{)}}_3{\mathrm{Sb}}_5$ are shown in (c) and (f), respectively. The markers in (c) and (f) indicate the substitution levels in our XRD measurements.
\label{fig_phase}}
\end{figure}

Figure\,\ref{fig_phase} shows the resistivity measurements of our Cs(V$_{1-x}M_{x}$)$_3$Sb$_5$ ($M$ = Nb, Ta) single crystals.
The kink in resistivity versus temperature curve indicates the CDW phase transition, which is more pronounced in the d$\rho$/dT curve (Fig.\,\ref{fig_phase} (a, d)). CDW is suppressed by substituting V with Ta and Nb, resulting in a decreased CDW transition temperature $T_{\mathrm{CDW}}$ and reduced sharpness of peak at $T_{\mathrm{CDW}}$ in the d$\rho$/dT curve. 
However, the superconducting transition temperature $T_{\mathrm{c}}$ shows a non-monotonic increase with substitution (Fig.\,\ref{fig_phase} (b) and (e)). For Ta-substituted CsV$_3$Sb$_5$, $T_{\mathrm{c}}$, determined by the midpoint of the resistivity drop, is enhanced from 2.8 K in the pristine sample to 3.5 K in x = 0.04, and then remains almost the same until x = 0.067. At x = 0.1, where the CDW phase is completely suppressed, $T_{\mathrm{c}}$ increases to 4.7 K (Fig.\,\ref{fig_phase} (b)). 
In Nb-substituted ${\mathrm{CsV}}_3{\mathrm{Sb}}_5$, $T_{\mathrm{CDW}}$ is suppressed from 94 K to 70 K with Nb substitution increasing to 0.068, while $T_{\mathrm{c}}$ is enhanced from 2.8 K in the pristine sample to $\sim$ 4K and then remains almost the same in the substitution range from 0.021 to 0.068, as shown in Fig.\,\ref{fig_phase} (d) and (e). The schematic phase diagrams for ${\mathrm{Cs(V}}_{1-x}{\mathrm{Ta}}_{x}{\mathrm{)}}_3{\mathrm{Sb}}_5$ and ${\mathrm{Cs(V}}_{1-x}{\mathrm{Nb}}_{x}{\mathrm{)}}_3{\mathrm{Sb}}_5$ obtained from resistivity measurement results are shown in Fig.\,\ref{fig_phase} (c) and (f), respectively. However, the study of Nb substitution persists only up to x $\simeq$ 0.7 due to the solubility limit, where the CDW is not fully suppressed\,\cite{CVNS_phaseDiagram}.

\subsection{XRD measurements of CDWs at low temperature}

\begin{figure}[htbp]
\centering
\includegraphics[width=\columnwidth]{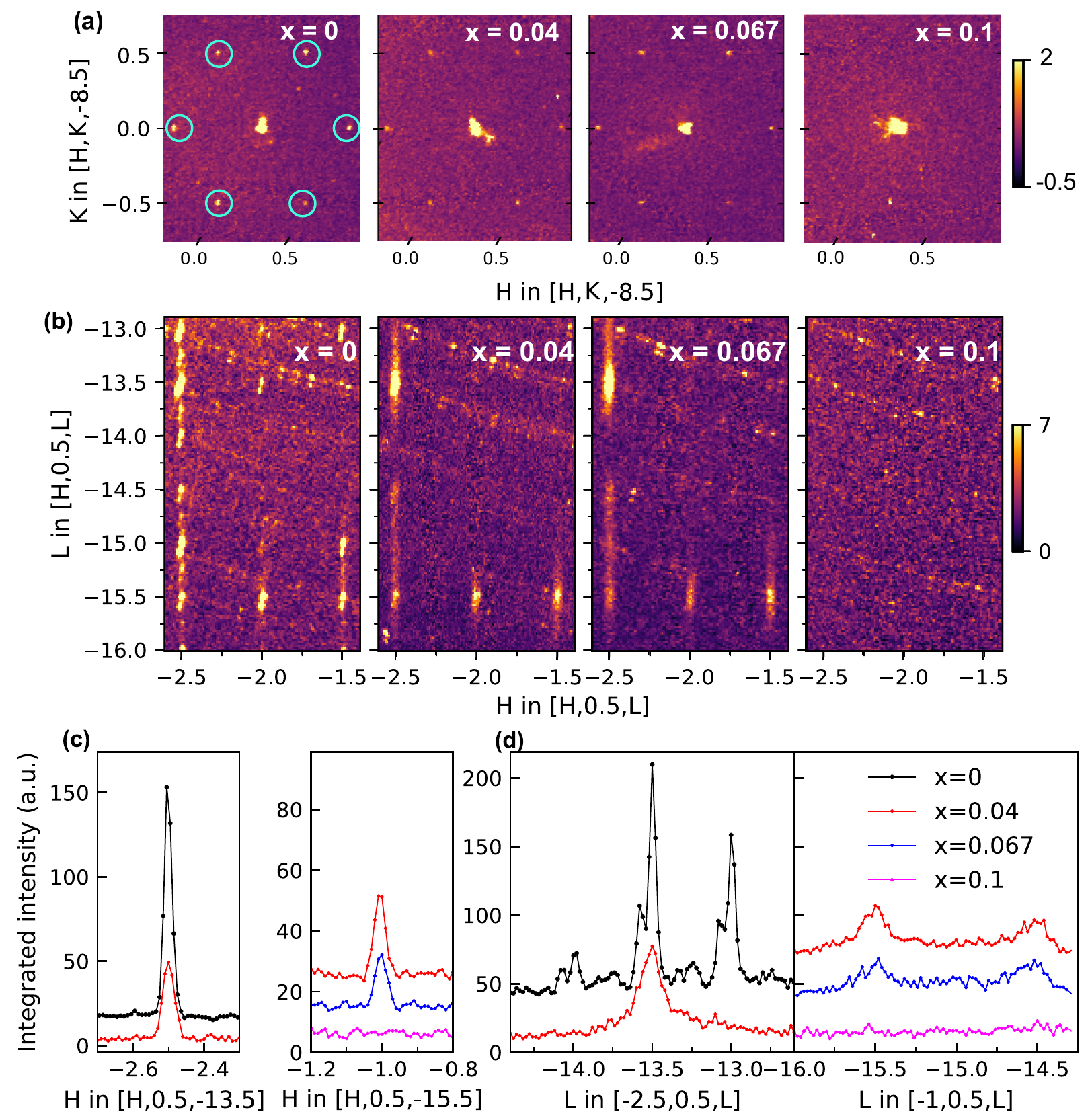}
\caption{XRD measurements for ${\mathrm{Cs(V}}_{1-x}{\mathrm{Ta}}_{x}{\mathrm{)}}_3{\mathrm{Sb}}_5$ with four different Ta substitution levels taken at 18 K using a Mo K$_{\alpha}$ x-ray source. (a) $(H, K)$ maps of reciprocal space at $L$ = -8.5. (b) $(H, L)$ maps of reciprocal space at $K$ = 0.5. The blue circles indicate the CDW diffraction points. (c) $H$-cuts of [-2.5, 0.5, -13.5] and [-1, 0.5, -15.5] are shown in the left and right panels, respectively. (d) $L$-cuts of [-2.5, 0.5, $L$] and [-1, 0.5, $L$] are shown in the left and right panels, respectively. An offset is added for clarity. The double peaks of $x$ = 0 originate from the small non-monochromatic x-ray source. The arc signals in colormaps come from beryllium domes. 
\label{figTa_Mo_source}}
\end{figure}

\begin{figure}[htbp]
\centering
\includegraphics[width=\columnwidth]{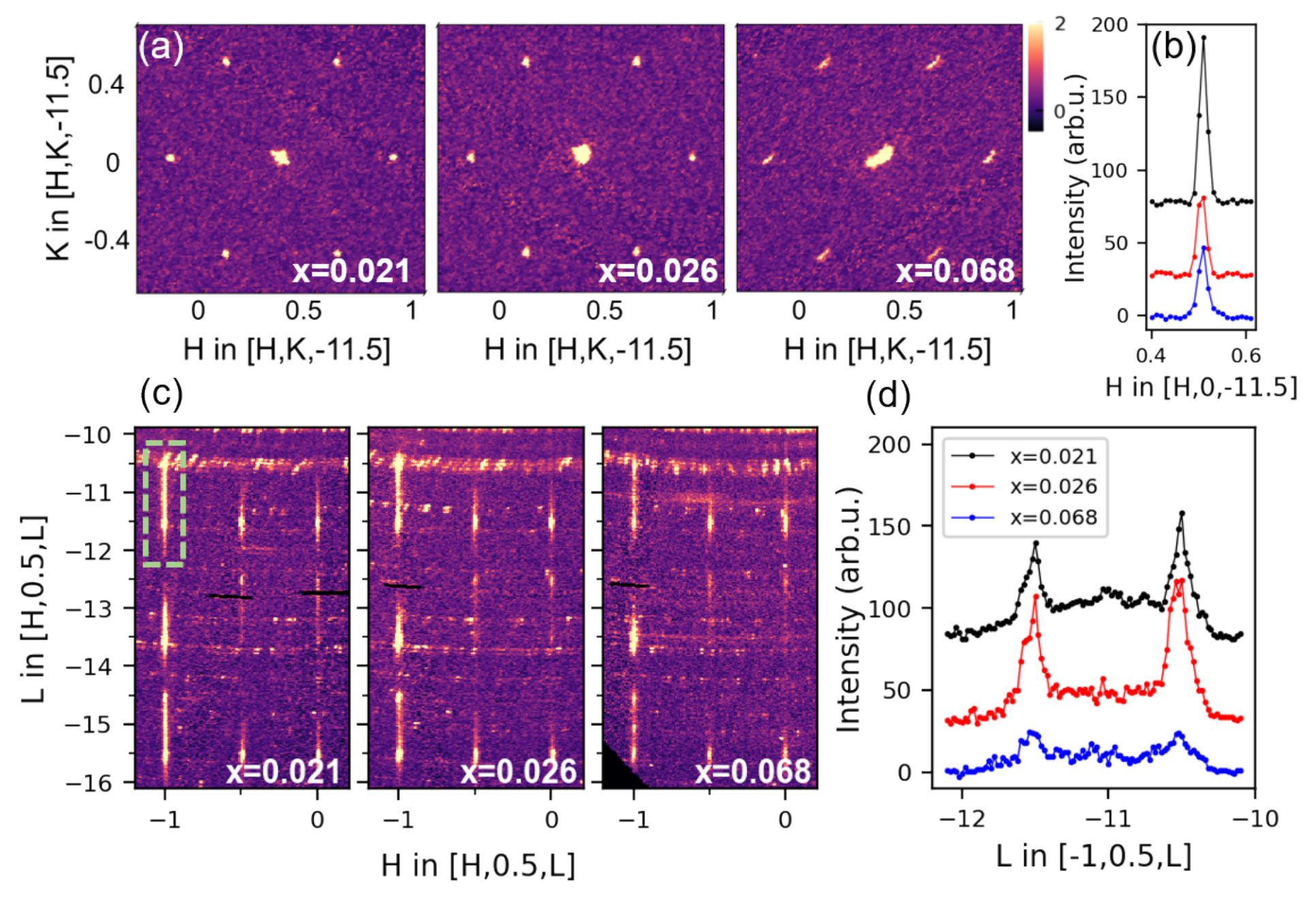}
\caption{(a) The $(H, K)$ maps of reciprocal space at $L$ = -11.5 for ${\mathrm{Cs(V}}_{1-x}{\mathrm{Nb}}_{x}{\mathrm{)}}_3{\mathrm{Sb}}_5$ with three different Nb substitution levels measured at 18 K using a Mo K$_\alpha$ x-ray source. (b) $H$-cuts of [H, 0, -11.5] at 18 K. Offset is added for clarity. (c) $(H, L)$ maps of reciprocal space at $K$ = 0.5. (d) $L$-cuts of [-1, 0.5, L] at 18 K. Offset is added for clarity. The arc signals in colormaps come from beryllium domes. 
\label{figNb}}
\end{figure}

In order to investigate the effects of substitution on the CDWs in ${\mathrm{Cs(V}}_{1-x}{\mathrm{Ta}}_{x}{\mathrm{)}}_3{\mathrm{Sb}}_5$ and ${\mathrm{Cs(V}}_{1-x}{\mathrm{Nb}}_{x}{\mathrm{)}}_3{\mathrm{Sb}}_5$, we conducted XRD measurements with different substitution levels. The samples were cooled from room temperature to 18 K at a rate of 2 K/min, and Mo K$_\alpha$ radiation was used as the x-ray source. Figure\,\ref{figTa_Mo_source} (a) shows the evolution of $(H, K)$ map at $L$ = -8.5 for ${\mathrm{Cs(V}}_{1-x}{\mathrm{Ta}}_{x}{\mathrm{)}}_3{\mathrm{Sb}}_5$. As V is substituted by Ta, the CDW diffraction points are weakened and eventually disappear at $x$ = 0.1. 
The changes in the CDW are more pronounced along the $L$ direction as shown in Fig.\,\ref{figTa_Mo_source} (b). In the $(H, L)$ map at $K$ = 0.5, the discrete CDW diffraction points with integer, half-integer, and quarter-integer $L$ values in pristine ${\mathrm{CsV}}_3{\mathrm{Sb}}_5$ become rod-like and are centered at half-integer $L$ in the substituted samples. This suggests that the CDW becomes quasi-two-dimensional (quasi-2D) as V is substituted by Ta. To better visualize the change of CDW peaks along in-plane and out-of-plane directions, we compared the $H$- and $L$-cuts at different substitution levels, as shown in Fig.\ref{figTa_Mo_source} (c)(d). It is observed that the widths of CDW peaks have broadened in both the in-plane and out-of-plane directions, as compared to the pristine sample. Figure\,\ref{figNb} shows the XRD measurements for ${\mathrm{Cs(V}}_{1-x}{\mathrm{Nb}}_{x}{\mathrm{)}}_3{\mathrm{Sb}}_5$ crystals. Similar to Ta substitution, the CDW diffraction signals of Nb-substituted ${\mathrm{CsV}}_3{\mathrm{Sb}}_5$ also become rod-like in the $(H, L)$ map at $K$ = 0.5, indicating a short-ranged correlation along the $L$-direction. These results suggest that substituting V with Ta/Nb hinders the establishment of long-ranged three-dimensional (3D) CDW correlations. 

\begin{figure}[htbp]
\centering
\includegraphics[width=\columnwidth]{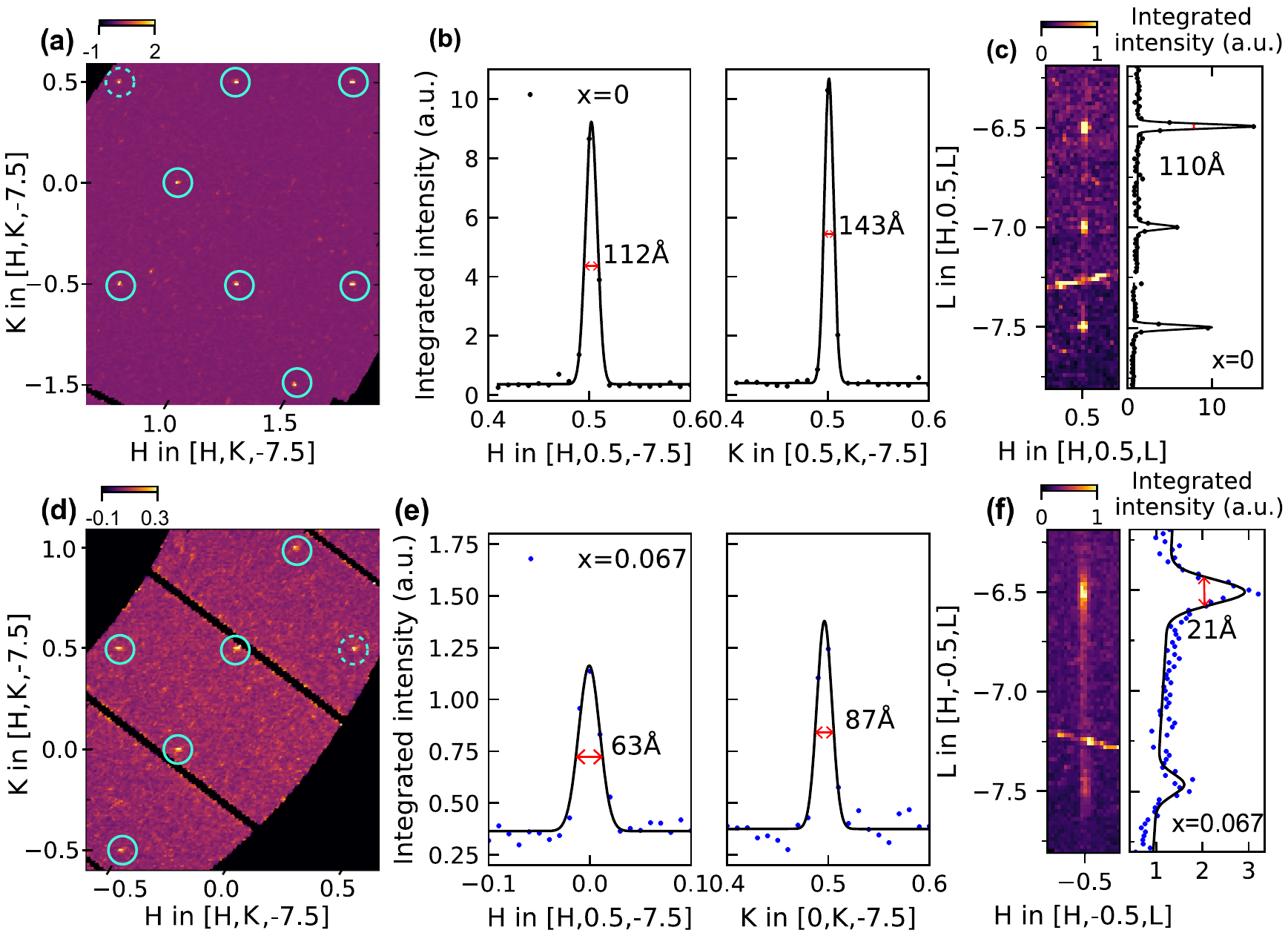}
\caption{XRD measurements collected at 18 K using a highly monochromatic
Cu K$_{\alpha 1}$ (8.048 keV) radiation x-ray source for  ${\mathrm{Cs(V}}_{1-x}{\mathrm{Ta}}_{x}{\mathrm{)}}_3{\mathrm{Sb}}_5$. (a)(d) $(H, K)$ maps of reciprocal space at $L$ = -7.5 for the pristine sample and $x$ = 0.067, respectively. The cyan circles indicate the CDW diffraction points. The $H$- and $K$-cuts of CDW diffraction peaks highlighted by the dashed cyan circle in (a)(d) are shown in (b) and (e), respectively. (c)(f) $(H, L)$ maps of reciprocal space for the pristine sample at $K$ = 0.5 and $x$ = 0.067 at $K$ = -0.5, respectively. The corresponding $L$-cut is shown in the right panel. Solid black lines are fits with Gaussian functions and a linear background. The correlation length, defined as 1/HWHM, is shown in the figure. 
\label{figTa_Cu_source}}
\end{figure}

To quantitatively reveal the correlation lengths with substitution, we performed high-resolution XRD measurements
on pristine CsV$_3$Sb$_5$ and Cs(V$_{1-x}$Ta$_{x}$)$_3$Sb$_5$ ($x$ = 0.067) single crystals using a highly monochromatic Cu K$_{\alpha1}$ (8.048 keV) radiation x-ray source at 18 K. The upper panel and lower panel of Fig.\,\ref{figTa_Cu_source} show the results for the pristine and substituted samples, respectively. We fitted the CDW profiles along both in-plane and out-of-plane directions using Gaussian functions and extracted the correlation length $\xi$, defined as 1/HWHM (half width of half maximum). The in-plane correlation lengths along $H$- and $K$- directions reduced from $\sim$ 112 ${\mathrm{\AA}}$ and $\sim$ 143 ${\mathrm{\AA}}$ in the pristine sample to $\sim$ 63 ${\mathrm{\AA}}$ and $\sim$ 87 ${\mathrm{\AA}}$ in the x = 0.067 sample [see Fig.\,\ref{figTa_Cu_source}(b)(e) for comparison]. It is worth noting that the correlation lengths in the pristine samples are limited by resolution, whereas this is not the case for the substituted samples. We notice that the in-plane CDW reflections are anisotropic, as the correlation length along the $K$-direction is longer than the $H$-direction in both pristine and substituted samples. The estimated anisotropy, defined as ($\xi_{K}$-$\xi_{H}$)/($\xi_{K}$+$\xi_{H}$), increases from 12$\%$ in the pristine sample to 16$\%$ in the x = 0.067 sample. Whether this anisotropy may relate to electron nematicity needs further investigation. Intriguingly, the correlation length along $L$ direction significantly reduces from $\sim$ 110 ${\mathrm{\AA}}$ in the pristine sample to $\sim$ 21 ${\mathrm{\AA}}$ in Cs(V$_{1-x}$Ta$_{x}$)$_3$Sb$_5$ ($x$ = 0.067), indicating that only the neighboring inter-layers are correlated.

\subsection{Temperature evolution of CDW in substituted samples}
\begin{figure}[htbp]
\centering
\includegraphics[width=\columnwidth]{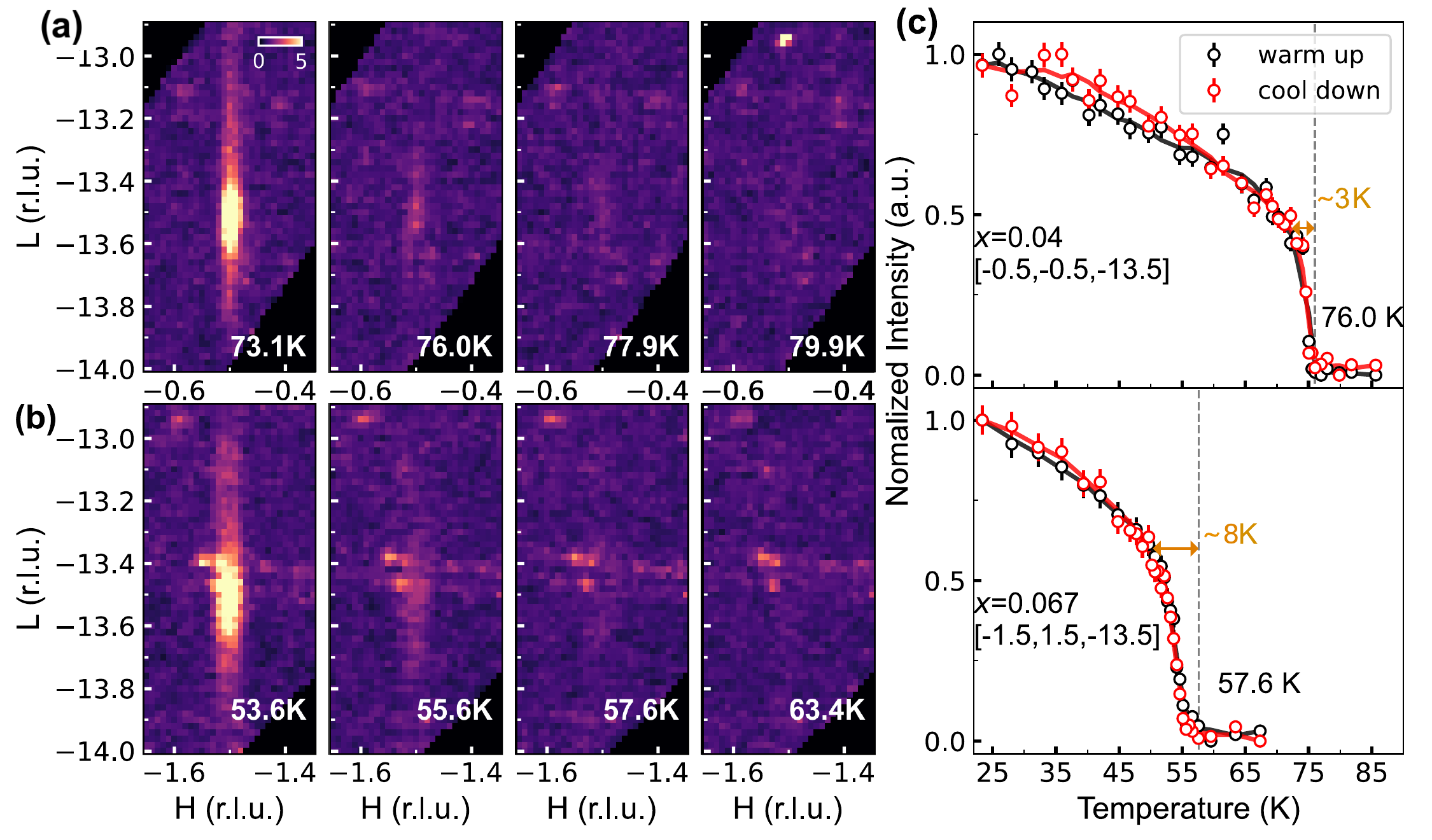}
\caption{Temperature dependence of the CDW in Ta-substituted samples. (a) The CDW profiles in $x$ = 0.04 measured at 73.1 K, 76 K, 77.9 K, and 79.9 K. (b) The CDW profiles in $x$ = 0.067 measured at 53.6 K, 55.6 K, 57.6 K and 63.4 K. (c) Temperature dependence of CDW in two Ta-substituted samples. The transition temperature and the transition width are indicated in the figure. The temperature evolution curves overlap well during the warming and cooling processes.
\label{figTaTdep}}
\end{figure}

\begin{figure}[htbp]
\centering\includegraphics[width = \columnwidth]{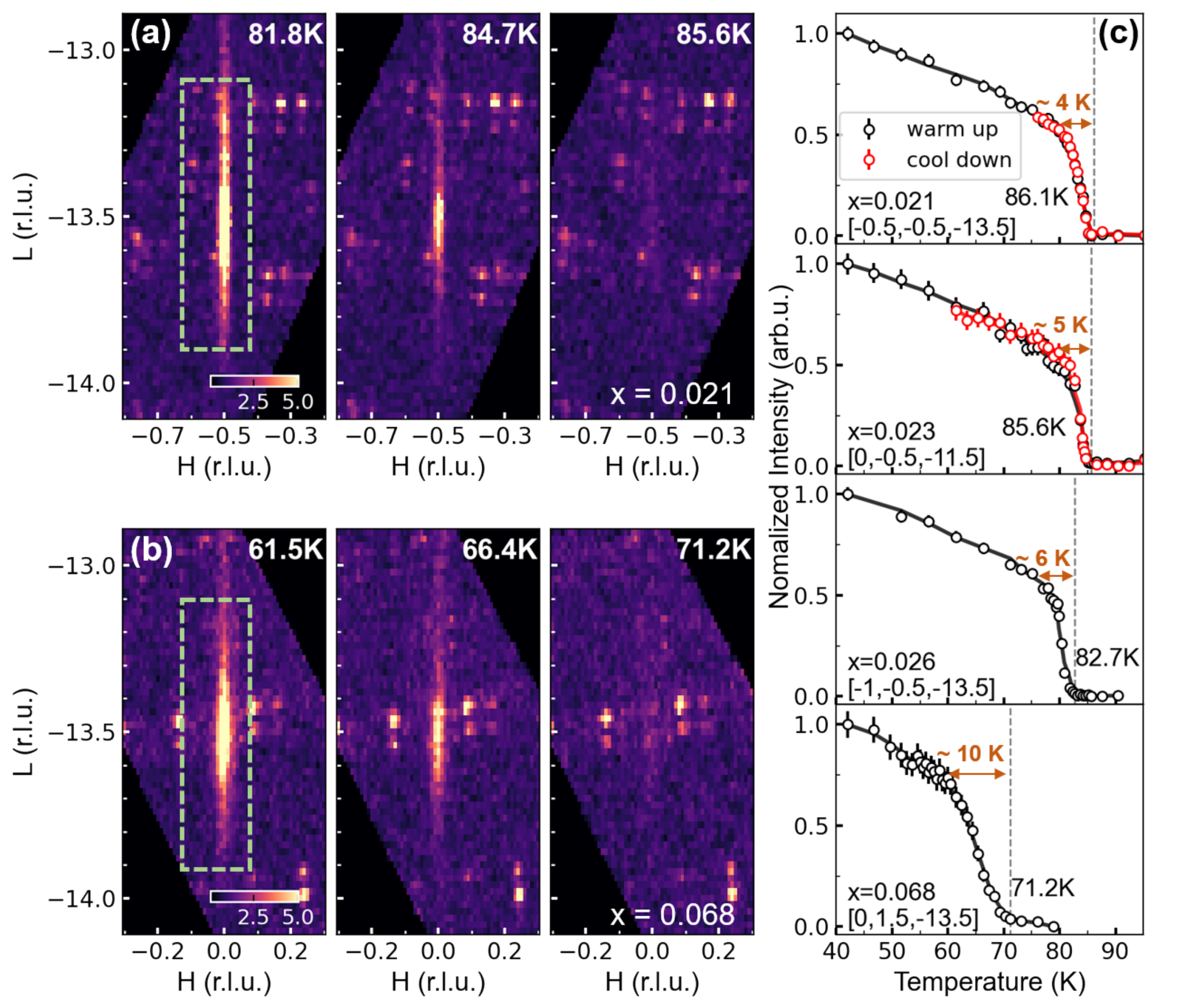}
\caption{\label{fig:NbTdep}
Temperature dependence of the CDW in Nb-substituted samples. (a) The CDW profiles in $x$ = 0.021 measured at 81.8 K, 84.7 K, and 85.6 K. (b) The CDW profiles in $x$ = 0.068 measured at 61.5 K, 66.4 K, and 71.2 K. (c) Temperature dependence of CDW in four Nb-substituted samples. The transition temperature and the transition width are indicated in the figure. The temperature evolution curves also show good agreement during the warming and cooling processes.}
\end{figure}

To investigate the temperature evolution of the CDW phase in Ta- and Nb-substituted ${\mathrm{CsV}}_3{\mathrm{Sb}}_5$ samples, we conducted temperature-dependent XRD measurements using a Mo
K$_\alpha$ radiation x-ray source. We cooled the samples from room temperature to 18 K at a rate of 2 K/min, then performed warming and cooling measurements. The resulting data for ${\mathrm{Cs(V}}_{1-x}{\mathrm{Ta}}_{x}{\mathrm{)}}_3{\mathrm{Sb}}_5$ and ${\mathrm{Cs(V}}_{1-x}{\mathrm{Nb}}_{x}{\mathrm{)}}_3{\mathrm{Sb}}_5$ are presented in Fig.\,\ref{figTaTdep} and Fig.\,\ref{fig:NbTdep}, respectively. Each data point is the average of six repeated measurements at each temperature to improve statistical accuracy, which took about one hour per temperature. The normalized intensity was obtained by integrating the entire CDW rods and normalizing the low-temperature signal.

We observed no thermal hysteresis behavior in the CDW diffraction signals in any of our substituted ${\mathrm{CsV}}_3{\mathrm{Sb}}_5$ samples. This is in sharp contrast to the pristine sample, where the 2 $\times$ 2 $\times$ 4 CDW undergoes a transition at 93 K in the warming process and 89 K in the cooling process, exhibiting a 4 K thermal hysteresis behavior\,\cite{xq}. The lack of thermal hysteresis behavior of CDW diffraction signals in Cs(V$_{1-x}M_{x}$)$_3$Sb$_5$ ($M$ = Nb, Ta) indicates the destruction of the 2$\times$2$\times$4 CDW phase. 
Consistent with our resistivity measurements, we find that the CDW transition temperature $T_{\mathrm{CDW}}$ decreases with substitution. In the pristine sample, $T_{\mathrm{CDW}}$ was 94 K, whereas it was 76 K and 57.6 K in the x = 0.04 and 0.067 Ta-substituted samples, respectively. There was no CDW observed in the x = 0.1 Ta-substituted sample. For the x = 0.04 Ta-substituted sample, we found that $T_{\mathrm{CDW}}$ was 76 K based on the CDW diffraction intensity versus temperature curve and resistivity measurements. However, there were some residual weak CDW diffraction signals present until 79.9 K [see Fig.\,\ref{figTaTdep}(a)]. This cannot be attributed to temperature fluctuations in our instrument, as our cryostat has a temperature stability of $\pm$ 30 mK. Nonetheless, this residual CDW signal was not universal among all substituted samples, and its origin remains unclear.

For Nb-substituted samples, we determined $T_{\mathrm{CDW}}$ to be 86.1 K, 85.6 K, 82.7 K, and 71.2 K for the x = 0.021, 0.023, 0.026, and 0.068 samples, respectively.
Interestingly, we observed that the CDW transition broadened with increasing substitution levels, as shown in Fig.\,\ref{figTaTdep} (c) and Fig.\,\ref{fig:NbTdep} (c). The CDW transition in the pristine sample is very sharp, with a transition width within 2 K\,\cite{xq}. In contrast, the CDW transition width broadens to 8 K in the Cs(V$_{1-x}$Ta$_{x}$)$_3$Sb$_5$ ($x$ = 0.067) sample and 10 K in the Cs(V$_{1-x}$Nb$_{x}$)$_3$Sb$_5$ ($x$ = 0.068) sample. 
These results suggest that the substitution destroys the formation of long-ranged CDW orders.

\section{Discussion}
Our results indicate that the CDW phase of ${\mathrm{Cs(V}}_{1-x}M_{x}{\mathrm{)}}_3{\mathrm{Sb}}_5$ ($M$ = Nb, Ta) is suppressed not only by a decrease in the CDW transition temperature but also by a reduction in the CDW correlation lengths, particularly along the $c$-direction. Specifically, the 2 $\times$ 2 $\times$ 4 CDW phase disappears with slight substitution and the remaining 2 $\times$ 2 $\times$ 2 CDW phase becomes quasi-2D with a correlation length of only about 21 ${\mathrm{\AA}}$, i.e., 2 unit-cells along the $c$ direction. In pristine CsV$_3$Sb$_5$, the CDW phases arise from inter-layer stacking of layers with the same in-plane CDW modulation\,\cite{nematic.CDW.nat,xq,TRSB.NematicDomain.natphy}. Although the phase shift between adjacent Kagome layers persists in substituted samples, the random stacking along the $c$ direction results in rod-like signals in the XRD pattern of the substituted samples. Similarly, the rod-like signals in Y$_5$Ru$_2$O$_{12}$ are suggested to originate from stacking faults\,\cite{Y5Ru2O12.stackingFault}.

Rod-like CDW signals have also been observed in Sn-doped CsV$_3$Sb$_5$, which are found centered at both half-integer and integer $L$ values\,\cite{CVSSn.xrd}. However, in our study, we find CDW diffraction peaks at integer $L$ to be invisible, likely due to their weak intensities. 
Some CDW diffraction signals centered at half-integer $L$ are asymmetric, leading us to speculate that the diffraction peaks at integer-$L$ are hidden in the broadened CDW reflection signals centered at half-integer $L$. Compared to Sn-doped ${\mathrm{CsV}}_3{\mathrm{Sb}}_{5}$\,\cite{CVSSn.xrd}, the CDW correlation lengths are shorter in both in-plane and out-of-plane directions in the substituted ${\mathrm{Cs(V}}_{1-x}M_{x}{\mathrm{)}}_3{\mathrm{Sb}}_5$ ($M$ = Nb, Ta). For example, in CsV$_3$Sb$_{5-x}$Sn$_x$, the CDW correlation length along the $H$-direction and $L$-direction is 367 ${\mathrm{\AA}}$ and 70 ${\mathrm{\AA}}$, respectively. The CDW correlation length along $H$-direction and $L$-direction in Cs(V$_{1-x}$Ta$_x$)$_3$Sb$_5$ (x = 0.067) is reduced to 63 ${\mathrm{\AA}}$ and 21 ${\mathrm{\AA}}$, respectively. 
Previous studies indicate that substitution of Sb primarily affects the Sb-derived electron-like band at $\Gamma$ point\,\cite{doping.2sc}, whereas substitution of V by Nb enlarges the electron pocket at $\Gamma$ point and lifts the vHSs at M point up and above Fermi level\,\cite{CVNS.ARPES.PRL,CVNS_phaseDiagram}. Therefore, the change of the electronic band at the M point degrades the nesting condition and suppresses the CDW. Additionally, substitution of V by Nb/Ta introduces disorder\,\cite{optical.transport.CVNS}, which further suppresses the CDW phase by reducing CDW correlation length\,\cite{PNAS.cuprate.disorder,npj.TaSe2-xSx.disorder,1983.prb.disorder}.

We then discuss the phase diagram of ${\mathrm{Cs(V}}_{1-x}M_{x}{\mathrm{)}}_3{\mathrm{Sb}}_5$ ($M$ = Nb, Ta) as shown in Fig.\,\ref{fig_phase}. While the CDW transition is rapidly suppressed with substitution, we observe that the increase in $T_{\mathrm{c}}$ becomes less efficient after the initial enhancement until the CDW phase is fully suppressed. 
Our XRD results demonstrate that substituting V by Ta/Nb causes the 2 $\times$ 2 $\times$ 4 CDW phase in CsV$_3$Sb$_5$ to disappear. 
Given the absence of the 2 $\times$ 2 $\times$ 4 CDW phase\,\cite{xq,sdh2021Brenden,stevern.xrd.AVS} and the unusual relationship between CDW and superconductivity are absent in KV$_3$Sb$_5$ and RbV$_3$Sb$_5$\,\cite{ChenKY.pressure.PRL,F.H.Yu.pressure.nc,doping.2sc,review.natphy,KVSSn.RVSSn.prm}, we speculate that the initial increase in $T_{\mathrm{c}}$ results from the destruction of the 2 $\times$ 2 $\times$ 4 CDW phase.
Subsequently, $T_{\mathrm{c}}$ remains almost unchanged as the Ta/Nb substitution level increases, possibly due to a complicated interplay among 2 $\times$ 2 $\times$ 2 CDW, electronic nematicity\,\cite{nematic.CDW.nat,wu2023unidirectional.Ti} and superconductivity. Only upon the complete disappearance of the CDW phase do we observe a second enhancement of $T_{\mathrm{c}}$.

We would like to note that we did not observe incommensurate quasi-1D electron correlations in ${\mathrm{Cs(V}}_{1-x}M_{x}{\mathrm{)}}_3{\mathrm{Sb}}_5$ ($M$ = Nb, Ta), as found in ${\mathrm{CsV}}_3{\mathrm{Sb}}_{5-x}$Sn$_x$\,\cite{CVSSn.xrd}. This difference could be related to the distinct phase diagrams of the two systems. In ${\mathrm{CsV}}_3{\mathrm{Sb}}_{5-x}$Sn$_x$, incommensurate quasi-1D electron correlations were observed in the region between the double peaks of $T_{\mathrm{c}}$. Similarly, a stripe-like CDW order was revealed by nuclear magnetic resonance measurements in pressurized ${\mathrm{CsV}}_3{\mathrm{Sb}}_5$\,\cite{cxh.nat.stripeCDW}. This stripe order is argued to be responsible for the suppression of $T_{\mathrm{c}}$ between double peaks\,\cite{cxh.nat.stripeCDW}, where the superconducting transition is broad\,\cite{F.H.Yu.pressure.nc,ChenKY.pressure.PRL}. However, the superconducting transition is sharp in both Sn-doped or Ta/Nb-substituted ${\mathrm{CsV}}_3{\mathrm{Sb}}_5$ samples. On the other hand, the incommensurate quasi-1D electron correlations observed in ${\mathrm{CsV}}_3{\mathrm{Sb}}_{5-x}$Sn$_x$ are likely related to the distribution of Sb atoms, which have been shown to be involved in CDW formation through resonant x-ray scattering study on CsV$_3$Sb$_5$\,\cite{li2021spatial}. The doping of Sb with Sn may prompt the emergence of quasi-1D electron correlations by inducing a reconstruction of Fermi pockets\,\cite{stripe.PDW}.
We also notice that the phase diagram of ${\mathrm{Cs(V}}_{1-x}M_{x}{\mathrm{)}}_3{\mathrm{Sb}}_5$ ($M$ = Nb, Ta) is different from that of ${\mathrm{Cs(V}}_{1-x}{\mathrm{Ti}}_{x}{\mathrm{)}}_3{\mathrm{Sb}}_5$, where $T_{\mathrm{c}}$ monotonically decreases with increasing Ti doping level\,\cite{Tidoped,liu2021Tidoping,wu2023unidirectional.Ti}. This is because holes are introduced in the system by substituting V with Ti, which is different from the isovalent substitution of V atoms with Ta/Nb. This leads to the shrinking of the electron pocket at the $\Gamma$ point\,\cite{Tidoped}, which is responsible for superconductivity\,\cite{doping.2sc,CVNS_phaseDiagram,CVNS.ARPES.PRL}, resulting in the decrease of $T_{\mathrm{c}}$ with Ti doping.

\section{Conclusions}
In summary, our comprehensive x-ray diffraction experiments reveal the evolution of CDW in ${\mathrm{Cs(V}}_{1-x}M_{x}{\mathrm{)}}_3{\mathrm{Sb}}_5$ ($M$ = Nb, Ta). Our results show that the long-range three-dimensional 2 $\times$ 2 $\times$ 4 and 2 $\times$ 2 $\times$ 2 CDW phases coexisting in the pristine crystal evolve into a short-range quasi-two-dimensional 2 $\times$ 2 $\times$ 2 CDW phase upon substituting V with Ta/Nb. We confirm that the 2 $\times$ 2 $\times$ 4 CDW phase is rapidly destroyed by even a slight substitution, as evidenced by the absence of hysteresis behavior in the substituted samples. The CDW correlation lengths in both in-plane and out-of-plane directions are reduced upon substitution. Specifically, in Cs(V$_{1-x}$Ta$_x$)$_3$Sb$_5$ (x = 0.067) sample, the correlation lengths along $H$- and $K$-directions are reduced to 63 ${\mathrm{\AA}}$ and 87 ${\mathrm{\AA}}$ respectively, while the correlation length along the out-of-plane direction is reduced to 21 ${\mathrm{\AA}}$ ($\sim$ 2c), indicating a two-dimensional CDW nature. Our experiments also show that we did not observe quasi-1D electron correlations in ${\mathrm{Cs(V}}_{1-x}M_{x}{\mathrm{)}}_3{\mathrm{Sb}}_5$ ($M$ = Nb, Ta) as those found in ${\mathrm{CsV}}_3{\mathrm{Sb}}_{5-x}$Sn$_x$\,\cite{CVSSn.xrd}, which demonstrates that substituting V sites is different from doping Sb sites in CsV$_3$Sb$_5$. Our results, therefore, uncover the evolution of CDW with isovalent substitution of V sites in CsV$_3$Sb$_5$-derived materials and provide new insights into the underlying connections between CDW and superconductivity.

\section{Acknowledgements}
Y.~Y.~P. is grateful for financial support from the Ministry of Science and Technology of China (2019YFA0308401 and 2021YFA1401903) and the National Natural Science Foundation of China (11974029). Z.W. acknowledges the financial support from the National Key Research and Development Program of China (Nos:2020YFA0308800, 2022YFA1403400), the National Science Foundation of China (NSFC) (grant No. 92065109), Beijing Natural Science Foundation (grant nos Z210006 and Z190006). Z.W. thanks the Analysis and Testing Center at BIT for assistance in facility support. Y. F. Guo acknowledges the financial support from the National Science Foundation of China (Grant No. 92065201).


\begin{thebibliography}{10}
\expandafter\ifx\csname url\endcsname\relax
  \def\url#1{\texttt{#1}}\fi
\expandafter\ifx\csname urlprefix\endcsname\relax\def\urlprefix{URL }\fi
\providecommand{\bibinfo}[2]{#2}
\providecommand{\eprint}[2][]{\url{#2}}

\bibitem{S.Yan2011science}
\bibinfo{author}{Yan, S.}, \bibinfo{author}{Huse, D.~A.} \&
  \bibinfo{author}{White, S.~R.}
\newblock \bibinfo{title}{{Spin-liquid ground state of the $S$ = 1/2 Kagome
  Heisenberg antiferromagnet}}.
\newblock \emph{\bibinfo{journal}{Science}} \textbf{\bibinfo{volume}{332}},
  \bibinfo{pages}{1173--1176} (\bibinfo{year}{2011}).

\bibitem{2016RevModPhys}
\bibinfo{author}{Norman, M.~R.}
\newblock \bibinfo{title}{{Colloquium: Herbertsmithite and the search for the
  quantum spin liquid}}.
\newblock \emph{\bibinfo{journal}{Rev. Mod. Phys.}}
  \textbf{\bibinfo{volume}{88}}, \bibinfo{pages}{041002}
  (\bibinfo{year}{2016}).

\bibitem{2017RevModPhys}
\bibinfo{author}{Zhou, Y.}, \bibinfo{author}{Kanoda, K.} \&
  \bibinfo{author}{Ng, T.-K.}
\newblock \bibinfo{title}{Quantum spin liquid states}.
\newblock \emph{\bibinfo{journal}{Rev. Mod. Phys.}}
  \textbf{\bibinfo{volume}{89}}, \bibinfo{pages}{025003}
  (\bibinfo{year}{2017}).

\bibitem{chiralSDW}
\bibinfo{author}{Yu, S.-L.} \& \bibinfo{author}{Li, J.-X.}
\newblock \bibinfo{title}{{Chiral superconducting phase and chiral
  spin-density-wave phase in a Hubbard model on the kagome lattice}}.
\newblock \emph{\bibinfo{journal}{Phys. Rev. B}} \textbf{\bibinfo{volume}{85}},
  \bibinfo{pages}{144402} (\bibinfo{year}{2012}).

\bibitem{instabilities.PRL}
\bibinfo{author}{Kiesel, M.~L.}, \bibinfo{author}{Platt, C.} \&
  \bibinfo{author}{Thomale, R.}
\newblock \bibinfo{title}{Unconventional fermi surface instabilities in the
  kagome hubbard model}.
\newblock \emph{\bibinfo{journal}{Phys. Rev. Lett.}}
  \textbf{\bibinfo{volume}{110}}, \bibinfo{pages}{126405}
  (\bibinfo{year}{2013}).

\bibitem{instabilities.prb}
\bibinfo{author}{Wang, W.-S.}, \bibinfo{author}{Li, Z.-Z.},
  \bibinfo{author}{Xiang, Y.-Y.} \& \bibinfo{author}{Wang, Q.-H.}
\newblock \bibinfo{title}{Competing electronic orders on kagome lattices at van
  hove filling}.
\newblock \emph{\bibinfo{journal}{Phys. Rev. B}} \textbf{\bibinfo{volume}{87}},
  \bibinfo{pages}{115135} (\bibinfo{year}{2013}).

\bibitem{Brenden.prl}
\bibinfo{author}{Ortiz, B.~R.} \emph{et~al.}
\newblock \bibinfo{title}{{$\mathrm{Cs}{\mathrm{V}}_{3}{\mathrm{Sb}}_{5}$: A
  ${\mathrm{Z}}_{2}$ topological Kagome metal with a superconducting ground
  state}}.
\newblock \emph{\bibinfo{journal}{Phys. Rev. Lett.}}
  \textbf{\bibinfo{volume}{125}}, \bibinfo{pages}{247002}
  (\bibinfo{year}{2020}).

\bibitem{prm2021sc.KVS}
\bibinfo{author}{Ortiz, B.~R.} \emph{et~al.}
\newblock \bibinfo{title}{{Superconductivity in the ${\mathrm{Z}}_{2}$ Kagome
  metal ${\mathrm{KV}}_{3}{\mathrm{Sb}}_{5}$}}.
\newblock \emph{\bibinfo{journal}{Phys. Rev. Mater.}}
  \textbf{\bibinfo{volume}{5}}, \bibinfo{pages}{034801} (\bibinfo{year}{2021}).

\bibitem{cpl2021RVS.sc}
\bibinfo{author}{Yin, Q.} \emph{et~al.}
\newblock \bibinfo{title}{{Superconductivity and normal-state properties of
  Kagome metal ${\mathrm{RbV}}_{3}{\mathrm{Sb}}_{5}$ single crystals}}.
\newblock \emph{\bibinfo{journal}{Chinese Physics Letters}}
  \textbf{\bibinfo{volume}{38}}, \bibinfo{pages}{037403}
  (\bibinfo{year}{2021}).

\bibitem{jiangyx}
\bibinfo{author}{Jiang, Y.-X.} \emph{et~al.}
\newblock \bibinfo{title}{{Unconventional chiral charge order in kagome
  superconductor ${\mathrm{KV}}_{3}{\mathrm{Sb}}_{5}$}}.
\newblock \emph{\bibinfo{journal}{Nature Materials}}
  \textbf{\bibinfo{volume}{20}}, \bibinfo{pages}{1353--1357}
  (\bibinfo{year}{2021}).

\bibitem{AHE.KVS}
\bibinfo{author}{Yang, S.-Y.} \emph{et~al.}
\newblock \bibinfo{title}{{Giant, unconventional anomalous Hall effect in the
  metallic frustrated magnet candidate,
  $\mathrm{K}{\mathrm{V}}_{3}{\mathrm{Sb}}_{5}$}}.
\newblock \emph{\bibinfo{journal}{Science Advances}}
  \textbf{\bibinfo{volume}{6}}, \bibinfo{pages}{eabb6003}
  (\bibinfo{year}{2020}).

\bibitem{AHE.CVS}
\bibinfo{author}{Yu, F.~H.} \emph{et~al.}
\newblock \bibinfo{title}{Concurrence of anomalous hall effect and charge
  density wave in a superconducting topological kagome metal}.
\newblock \emph{\bibinfo{journal}{Phys. Rev. B}}
  \textbf{\bibinfo{volume}{104}}, \bibinfo{pages}{L041103}
  (\bibinfo{year}{2021}).

\bibitem{umR.kenny}
\bibinfo{author}{Kenney, E.~M.}, \bibinfo{author}{Ortiz, B.~R.},
  \bibinfo{author}{Wang, C.}, \bibinfo{author}{Wilson, S.~D.} \&
  \bibinfo{author}{Graf, M.~J.}
\newblock \bibinfo{title}{{Absence of local moments in the kagome metal
  ${\mathrm{KV}}_{3}{\mathrm{Sb}}_{5}$ as determined by muon spin
  spectroscopy.}}
\newblock \emph{\bibinfo{journal}{Journal of Physics: Condensed Matter}}
  \textbf{\bibinfo{volume}{33}}, \bibinfo{pages}{235801}
  (\bibinfo{year}{2021}).

\bibitem{Tidoped}
\bibinfo{author}{Yang, H.} \emph{et~al.}
\newblock \bibinfo{title}{{Titanium doped kagome superconductor
  CsV$_{3-x}$Ti$_x$Sb$_5$ and two distinct phases}}.
\newblock \emph{\bibinfo{journal}{Science Bulletin}}
  \textbf{\bibinfo{volume}{67}}, \bibinfo{pages}{2176--2185}
  (\bibinfo{year}{2022}).

\bibitem{CVNS_phaseDiagram}
\bibinfo{author}{Li, Y.} \emph{et~al.}
\newblock \bibinfo{title}{{Tuning the competition between superconductivity and
  charge order in the kagome superconductor
  $\mathrm{Cs}{({\mathrm{V}}_{1\text{\ensuremath{-}}x}{\mathrm{Nb}}_{x})}_{3}{\mathrm{Sb}}_{5}$}}.
\newblock \emph{\bibinfo{journal}{Phys. Rev. B}}
  \textbf{\bibinfo{volume}{105}}, \bibinfo{pages}{L180507}
  (\bibinfo{year}{2022}).

\bibitem{NLWang.TRS}
\bibinfo{author}{Wu, Q.} \emph{et~al.}
\newblock \bibinfo{title}{{Simultaneous formation of two-fold rotation symmetry
  with charge order in the kagome superconductor
  ${\mathrm{CsV}}_{3}{\mathrm{Sb}}_{5}$ by optical polarization rotation
  measurement}}.
\newblock \emph{\bibinfo{journal}{Phys. Rev. B}}
  \textbf{\bibinfo{volume}{106}}, \bibinfo{pages}{205109}
  (\bibinfo{year}{2022}).

\bibitem{TRSB.NematicDomain.natphy}
\bibinfo{author}{Xu, Y.} \emph{et~al.}
\newblock \bibinfo{title}{{Three-state nematicity and magneto-optical Kerr
  effect in the charge density waves in kagome superconductors}}.
\newblock \emph{\bibinfo{journal}{Nature Physics}}
  \textbf{\bibinfo{volume}{18}}, \bibinfo{pages}{1470--1475}
  (\bibinfo{year}{2022}).

\bibitem{HongLi.natphy}
\bibinfo{author}{Li, H.} \emph{et~al.}
\newblock \bibinfo{title}{{Rotation symmetry breaking in the normal state of a
  kagome superconductor KV$_3$Sb$_5$}}.
\newblock \emph{\bibinfo{journal}{Nature Physics}}
  \textbf{\bibinfo{volume}{18}}, \bibinfo{pages}{265--270}
  (\bibinfo{year}{2022}).

\bibitem{2symmetry2021natcom}
\bibinfo{author}{Xiang, Y.} \emph{et~al.}
\newblock \bibinfo{title}{{Twofold symmetry of c-axis resistivity in
  topological kagome superconductor ${\mathrm{CsV}}_3{\mathrm{Sb}}_5$ with
  in-plane rotating magnetic field}}.
\newblock \emph{\bibinfo{journal}{Nature Communications}}
  \textbf{\bibinfo{volume}{12}}, \bibinfo{pages}{6727} (\bibinfo{year}{2021}).

\bibitem{nematic.CDW.nat}
\bibinfo{author}{Nie, L.} \emph{et~al.}
\newblock \bibinfo{title}{{Charge-density-wave-driven electronic nematicity in
  a kagome superconductor}}.
\newblock \emph{\bibinfo{journal}{Nature}} \textbf{\bibinfo{volume}{604}},
  \bibinfo{pages}{59--64} (\bibinfo{year}{2022}).

\bibitem{miuSR.nat}
\bibinfo{author}{Mielke, C.} \emph{et~al.}
\newblock \bibinfo{title}{{Time-reversal symmetry-breaking charge order in a
  kagome superconductor}}.
\newblock \emph{\bibinfo{journal}{Nature}} \textbf{\bibinfo{volume}{602}},
  \bibinfo{pages}{245--250} (\bibinfo{year}{2022}).

\bibitem{xq}
\bibinfo{author}{Xiao, Q.} \emph{et~al.}
\newblock \bibinfo{title}{{Coexistence of multiple stacking charge density
  waves in kagome superconductor ${\mathrm{CsV}}_{3}{\mathrm{Sb}}_{5}$}}.
\newblock \emph{\bibinfo{journal}{Phys. Rev. Res.}}
  \textbf{\bibinfo{volume}{5}}, \bibinfo{pages}{L012032}
  (\bibinfo{year}{2023}).

\bibitem{F.H.Yu.pressure.nc}
\bibinfo{author}{Yu, F.~H.} \emph{et~al.}
\newblock \bibinfo{title}{{Unusual competition of superconductivity and
  charge-density-wave state in a compressed topological kagome metal}}.
\newblock \emph{\bibinfo{journal}{Nature Communications}}
  \textbf{\bibinfo{volume}{12}}, \bibinfo{pages}{3645} (\bibinfo{year}{2021}).

\bibitem{ChenKY.pressure.PRL}
\bibinfo{author}{Chen, K.~Y.} \emph{et~al.}
\newblock \bibinfo{title}{{Double superconducting dome and triple enhancement
  of ${T}_{c}$ in the Kagome superconductor
  ${\mathrm{CsV}}_{3}{\mathrm{Sb}}_{5}$ under high pressure}}.
\newblock \emph{\bibinfo{journal}{Phys. Rev. Lett.}}
  \textbf{\bibinfo{volume}{126}}, \bibinfo{pages}{247001}
  (\bibinfo{year}{2021}).

\bibitem{doping.2sc}
\bibinfo{author}{Oey, Y.~M.} \emph{et~al.}
\newblock \bibinfo{title}{{Fermi level tuning and double-dome superconductivity
  in the Kagome metal
  ${\mathrm{CsV}}_{3}{\mathrm{Sb}}_{5\ensuremath{-}x}{\mathrm{Sn}}_{x}$}}.
\newblock \emph{\bibinfo{journal}{Phys. Rev. Mater.}}
  \textbf{\bibinfo{volume}{6}}, \bibinfo{pages}{L041801}
  (\bibinfo{year}{2022}).

\bibitem{KVSSn.RVSSn.prm}
\bibinfo{author}{Oey, Y.~M.}, \bibinfo{author}{Kaboudvand, F.},
  \bibinfo{author}{Ortiz, B.~R.}, \bibinfo{author}{Seshadri, R.} \&
  \bibinfo{author}{Wilson, S.~D.}
\newblock \bibinfo{title}{{Tuning charge density wave order and
  superconductivity in the Kagome metals
  ${\mathrm{KV}}_{3}{\mathrm{Sb}}_{5\ensuremath{-}x}\phantom{\rule{0.16em}{0ex}}{\mathrm{Sn}}_{x}$
  and
  ${\mathrm{RbV}}_{3}{\mathrm{Sb}}_{5\ensuremath{-}x}\phantom{\rule{0.16em}{0ex}}{\mathrm{Sn}}_{x}$}}.
\newblock \emph{\bibinfo{journal}{Phys. Rev. Mater.}}
  \textbf{\bibinfo{volume}{6}}, \bibinfo{pages}{074802} (\bibinfo{year}{2022}).

\bibitem{review.natphy}
\bibinfo{author}{Neupert, T.}, \bibinfo{author}{Denner, M.~M.},
  \bibinfo{author}{Yin, J.-X.}, \bibinfo{author}{Thomale, R.} \&
  \bibinfo{author}{Hasan, M.~Z.}
\newblock \bibinfo{title}{{Charge order and superconductivity in kagome
  materials}}.
\newblock \emph{\bibinfo{journal}{Nature Physics}}
  \textbf{\bibinfo{volume}{18}}, \bibinfo{pages}{137--143}
  (\bibinfo{year}{2022}).

\bibitem{CVTS.CVNS.ARPES}
\bibinfo{author}{Zhong, Y.} \emph{et~al.}
\newblock \bibinfo{title}{{Nodeless electron pairing in
  ${\mathrm{CsV}}_3{\mathrm{Sb}}_5$-derived kagome superconductors}}.
\newblock \eprint{arXiv:2303.00875}.

\bibitem{CVNS.ARPES.PRL}
\bibinfo{author}{Kato, T.} \emph{et~al.}
\newblock \bibinfo{title}{{Fermiology and origin of ${T}_{c}$ enhancement in a
  Kagome superconductor
  $\mathrm{Cs}({\mathrm{V}}_{1\ensuremath{-}x}{\mathrm{Nb}}_{x}{)}_{3}{\mathrm{Sb}}_{5}$}}.
\newblock \emph{\bibinfo{journal}{Phys. Rev. Lett.}}
  \textbf{\bibinfo{volume}{129}}, \bibinfo{pages}{206402}
  (\bibinfo{year}{2022}).

\bibitem{CVSSn.xrd}
\bibinfo{author}{Kautzsch, L.} \emph{et~al.}
\newblock \bibinfo{title}{{Incommensurate charge-stripe correlations in the
  kagome superconductor ${\mathrm{CsV}}_3{\mathrm{Sb}}_{5-x}$Sn$_x$}}.
\newblock \eprint{arXiv:2207.10608}.

\bibitem{Y5Ru2O12.stackingFault}
\bibinfo{author}{Sanjeewa, L.~D.} \emph{et~al.}
\newblock \bibinfo{title}{{Stacking faults and short-range magnetic
  correlations in single crystal Y$_5$Ru$_2$O$_{12}$: a structure with
  Ru$^{+ 4.5}$ one-dimensional chains}}.
\newblock \emph{\bibinfo{journal}{physica status solidi (b)}}
  \textbf{\bibinfo{volume}{258}}, \bibinfo{pages}{2000197}
  (\bibinfo{year}{2020}).

\bibitem{optical.transport.CVNS}
\bibinfo{author}{Zhou, X.} \emph{et~al.}
\newblock \bibinfo{title}{{Effects of niobium doping on the charge density wave
  and electronic correlations in the kagome metal
  $\text{Cs}{({\text{V}}_{1\ensuremath{-}x}{\text{Nb}}_{x})}_{3}{\text{Sb}}_{5}$}}.
\newblock \emph{\bibinfo{journal}{Phys. Rev. B}}
  \textbf{\bibinfo{volume}{107}}, \bibinfo{pages}{125124}
  (\bibinfo{year}{2023}).

\bibitem{PNAS.cuprate.disorder}
\bibinfo{author}{Leroux, M.} \emph{et~al.}
\newblock \bibinfo{title}{{Disorder raises the critical temperature of a
  cuprate superconductor}}.
\newblock \emph{\bibinfo{journal}{Proceedings of the National Academy of
  Sciences}} \textbf{\bibinfo{volume}{116}}, \bibinfo{pages}{10691--10697}
  (\bibinfo{year}{2019}).

\bibitem{npj.TaSe2-xSx.disorder}
\bibinfo{author}{Li, L.} \emph{et~al.}
\newblock \bibinfo{title}{{Superconducting order from disorder in
  2H-TaSe$_{2-x}$S$_x$}}.
\newblock \emph{\bibinfo{journal}{npj Quantum Materials}}
  \textbf{\bibinfo{volume}{2}}, \bibinfo{pages}{11} (\bibinfo{year}{2017}).

\bibitem{1983.prb.disorder}
\bibinfo{author}{Mutka, H.}
\newblock \bibinfo{title}{{Superconductivity in irradiated charge-density-wave
  compounds $2H\ensuremath{-}\mathrm{Nb}{\mathrm{Se}}_{2}$,
  $2H\ensuremath{-}\mathrm{Ta}{\mathrm{S}}_{2}$, and
  $2H\ensuremath{-}\mathrm{Ta}{\mathrm{Se}}_{2}$}}.
\newblock \emph{\bibinfo{journal}{Phys. Rev. B}} \textbf{\bibinfo{volume}{28}},
  \bibinfo{pages}{2855--2858} (\bibinfo{year}{1983}).

\bibitem{sdh2021Brenden}
\bibinfo{author}{Ortiz, B.~R.} \emph{et~al.}
\newblock \bibinfo{title}{{Fermi surface mapping and the nature of
  charge-density-wave order in the Kagome superconductor
  ${\mathrm{CsV}}_{3}{\mathrm{Sb}}_{5}$}}.
\newblock \emph{\bibinfo{journal}{Phys. Rev. X}} \textbf{\bibinfo{volume}{11}},
  \bibinfo{pages}{041030} (\bibinfo{year}{2021}).

\bibitem{stevern.xrd.AVS}
\bibinfo{author}{Kautzsch, L.} \emph{et~al.}
\newblock \bibinfo{title}{{Structural evolution of the Kagome superconductors
  $A{\mathrm{V}}_{3}{\mathrm{Sb}}_{5}$ (A = K, Rb, and Cs) through charge
  density wave order}}.
\newblock \emph{\bibinfo{journal}{Phys. Rev. Mater.}}
  \textbf{\bibinfo{volume}{7}}, \bibinfo{pages}{024806} (\bibinfo{year}{2023}).

\bibitem{wu2023unidirectional.Ti}
\bibinfo{author}{Wu, P.} \emph{et~al.}
\newblock \bibinfo{title}{{Unidirectional electron-phonon coupling as a
  ``fingerprint'' of the nematic state in a kagome superconductor}}
  (\bibinfo{year}{2023}).
\newblock \eprint{arXiv:2302.05115}.

\bibitem{cxh.nat.stripeCDW}
\bibinfo{author}{Zheng, L.} \emph{et~al.}
\newblock \bibinfo{title}{{Emergent charge order in pressurized kagome
  superconductor $\mathrm{Cs}{\mathrm{V}}_{3}{\mathrm{Sb}}_{5}$}}.
\newblock \emph{\bibinfo{journal}{Nature}} \textbf{\bibinfo{volume}{611}},
  \bibinfo{pages}{682--687} (\bibinfo{year}{2022}).

\bibitem{li2021spatial}
\bibinfo{author}{Li, H.} \emph{et~al.}
\newblock \bibinfo{title}{{Discovery of conjoined charge density waves in the
  kagome superconductor ${\mathrm{CsV}}_3{\mathrm{Sb}}_5$}.}
\newblock \emph{\bibinfo{journal}{Nature Communications}}
  \textbf{\bibinfo{volume}{13}}, \bibinfo{pages}{6348} (\bibinfo{year}{2022}).

\bibitem{stripe.PDW}
\bibinfo{author}{Li, H.} \emph{et~al.}
\newblock \bibinfo{title}{{Small Fermi pockets intertwined with charge stripes
  and pair density wave order in a kagome superconductor}}.
\newblock \eprint{arXiv:2303.07254}.

\bibitem{liu2021Tidoping}
\bibinfo{author}{Liu, Y.} \emph{et~al.}
\newblock \bibinfo{title}{{Doping evolution of superconductivity, charge order
  and band topology in hole-doped topological kagome superconductors
  Cs(V$_{1-x}$Ti$_x$)$_3$Sb$_5$}}.
\newblock \eprint{arXiv:2110.12651}.

\end{thebibliography}
\end{document}